\documentclass[aps,prb,twocolumn,amsmath,superscriptaddress]{revtex4}

\usepackage{graphicx}

\usepackage{color}

\begin{document}

\title{Electron-nuclei spin dynamics in II-VI semiconductor quantum dots}

\author{C. Le Gall}
\affiliation{CEA-CNRS group "Nanophysique et
semiconducteurs", Institut N\'eel, CNRS \& Universit\'e
Joseph Fourier, BP 166, F-38042 Grenoble Cedex 9, France
}

\author{A. Brunetti}
\affiliation{CEA-CNRS group "Nanophysique et
semiconducteurs", Institut N\'eel, CNRS \& Universit\'e
Joseph Fourier, BP 166, F-38042 Grenoble Cedex 9, France
}

\author{H. Boukari}
\affiliation{CEA-CNRS group "Nanophysique et
semiconducteurs", Institut N\'eel, CNRS \& Universit\'e
Joseph Fourier, BP 166, F-38042 Grenoble Cedex 9, France
}

\author{L. Besombes}
\email{lucien.besombes@grenoble.cnrs.fr}
\affiliation{CEA-CNRS group "Nanophysique et
semiconducteurs", Institut N\'eel, CNRS \& Universit\'e
Joseph Fourier, BP 166, F-38042 Grenoble Cedex 9, France
}

\date{\today}

\begin{abstract}
We report on the dynamics of optically induced nuclear spin
polarization in individual CdTe/ZnTe quantum dots loaded
with one electron by modulation doping. The fine structure
of the hot trion (charged exciton $X^-$ with an electron in
the $P$-shell) is identified in photoluminescence
excitation spectra. A negative polarisation rate of the
photoluminescence, optical pumping of the resident electron
and the built-up of dynamic nuclear spin polarisation
(DNSP) are observed in time-resolved optical pumping
experiments when the quantum dot is excited at higher
energy than the hot trion triplet state. The time and
magnetic field dependence of the polarisation rate of the
$X^-$ emission allows to probe the dynamics of formation of
the DNSP in the optical pumping regime. We demonstrate
using time-resolved measurements that the creation of a
DNSP at B=0T efficiently prevents longitudinal spin
relaxation of the electron caused by fluctuations of the
nuclear spin bath. The DNSP is built in the microsecond
range at high excitation intensity. A relaxation time of
the DNSP in about 10 microseconds is observed at $B=0T$ and
significantly increases under a magnetic field of a few
milli-Tesla. We discuss mechanisms responsible for the fast
initialisation and relaxation of the diluted nuclear spins
in this system.
\end{abstract}

\maketitle

\section{Introduction}
The confinement of single electrons in semiconductor
quantum dots (QDs) and the control of their spin has been
motivated by perspectives of using the electron spin as the
ultimate solid state system to store and process quantum
information. In the commonly studied III-V semiconductor
QDs, the hyperfine interaction of the electron with the
fluctuating nuclear spins limits the time scale on which an
electron spin can be manipulated at low magnetic field. It
has been proposed that a full polarization of the nuclei
could cancel the decoherence of the electron induced by the
fluctuating hyperfine field \cite{Loss2003}. Alternatively,
the decoherence created by nuclei could be circumvented by
using isotopically purified II-VI materials \cite{Liu2007}
since Zn, Cd, Mg, O, Se and Te all have dominant isotopes
without nuclear spins. However, as highlighted in reference
3 and 4, the interaction between a confined electron in a
II-VI QD and the low density of nuclear spins I=1/2 in a QD
volume ten to one hundred times smaller than InAs/GaAs QDs
leads to some spin dynamics which is fundamentally
different from the one observed in III-V systems
\cite{Feng2007,Akimov2006}.

Due to the small QD size and low density of nuclear spins,
the electron-nuclei dynamics in II-VI QDs is ruled by a
large Knight field and significant nuclear spin
fluctuations despite a small Overhauser field.
Consequently, the nuclei-induced spin decoherence of the
electron is also an issue in II-VI QDs. However, the
built-up of a dynamic nuclear spin polarisation (DNSP) at
B=0T, can be much faster than the relaxation induced by the
dipole interaction between nuclear spins allowing the
creation of a strong non-equilibrium DNSP \cite{Feng2007}.
Under these conditions, decoherence of the electron should
be efficiently suppressed. Experimental study of the
electron-nuclear dynamics in II-VI QDs are few. Providing a
thorough experimental study of this system at a single dot
level is the aim of this work.

In this paper, we report on the dynamics of coupled
electron and nuclear spins polarization in individual
CdTe/ZnTe QDs with a resident electron introduced by
aluminium modulation doping. Here, in contrast to gated
structures where the number of resident charges is
controlled by an applied voltage, we use the characteristic
spectral feature of the charged exciton triplet state
observed in photoluminescence excitation spectra (PLE) to
identify QDs containing a single resident electron.
Non-resonant circularly polarized excitation of the
negatively charged exciton has been shown to lead to a
polarisation of the nuclear spins in both III-V
\cite{Eble2006} and II-VI QDs \cite{Akimov2006} and will be
used throughout this study to build-up and probe DNSP. In
II-VI QDs, the Overhauser shift is much smaller than the
photoluminescence (PL) linewidth and cannot be observed
directly: The nuclear field is detected through the
polarization rate of the resident electron which is
controlled by the nuclear spin fluctuations (NSF).

We give a description of the studied structures and
experimental techniques in section II of this paper. In
section III, we discuss nuclear spin-polarisation in II-VI
QDs, and derive orders of magnitude for the effective
hyperfine fields encountered in this system. Experimental
evidences of $X^-$ triplet states and details about the
mechanism of optical spin injection and build-up of
negative circular polarization in singly charged CdTe/ZnTe
QDs are presented in section IV. In section V we describe
how the polarization of nuclear spins influences spin
dynamics of the confined electron. The dynamics of nuclear
spin polarization is considered in section VI. Finally, in
section VII we present and discuss results of coupled
electron/nuclei spin decay in the absence of optical
excitation.

\section{Sample and experiment}

The sample used in this study is grown on a ZnTe substrate
and contains CdTe/ZnTe QDs. A 6.5- monolayer-thick CdTe
layer is deposited at $280\,^{\circ}\mathrm{C}$ by atomic
layer epitaxy on a ZnTe barrier grown by molecular beam
epitaxy at $360\,^{\circ}\mathrm{C}$. The dots are formed
by the high Tellurium deposition process described in
reference 7 and protected by a 100-nm-thick ZnTe top
barrier \cite{Wojnar2011}. A 20 nm thick Al doped ZnTe
layer is introduced 30 nm above the QDs leading to an
average negative charging of the QDs. The height of the QDs
core is about 2-3 nanometers and their diameter is 10 to 20
nm. We estimate an average QD volume of about 250 $nm^3$
containing $\approx 8000$ nuclei, $1200$ of which carry a
spin 1/2.

Optical addressing of individual QDs containing a single
electron is achieved using micro-spectroscopy techniques.
A high refractive index hemispherical solid immersion lens
is mounted on the bare surface of the sample to enhance
the spatial resolution and the collection efficiency of
single-dot emission in a low-temperature ($T=5K$) scanning
optical microscope \cite{LeGall2010}. Despite the quite
large QD density ($\approx 10^{10}cm^{-2}$) and the large
number of dots in the focal spot area, single QD
transitions can be identified by their spectral signatures.
A weak magnetic field of a few tens of milli-Tesla can be
applied in Voigt or Faraday configuration using permanent
magnets.

To investigate the mechanisms of spin injection, the QDs
are excited with energy tunable picosecond ($\approx 2ps$)
laser pulses from a frequency doubled optical parametric
oscillator with a repetition time of 13ns. A delay line can
be used to divide a single pulse into one {\em co} and one
{\em cross}-polarized pulses. Time-resolved experiments to
observe slower dynamics (optical pumping of electron and
nuclei) are performed using a modulated tunable continuous
wave (CW) dye laser. Laser pulses of controllable duration
and polarisation are created using Acousto-Optic Modulators
or an Electro-Optic modulator with rise times of about
10ns. The collected PL is dispersed by a $1m$ double
monochromator before being detected by a CCD camera or a
fast avalanche photodiode in conjunction with a
time-correlated photon-counting unit with an overall time
resolution of about 50 ps. In CW experiments, the
polarisation rate of the PL is measured using a
birefringent prism to separate the $\sigma+$ from the
$\sigma-$ component and to detect them at the same time on
different areas of the CCD camera.

\section{Nuclear spin polarization in II-VI quantum dots}

In a singly charged QD under the injection of spin
polarized electrons, a nuclear spin polarization builds up
by integration over many mutual spin flip-flops of the
confined electrons and the lattice nuclei. This nuclear
magnetic field modifies the coherent electron spin dynamics
and consequently the average polarization of the PL of the
$X^-$. The knowledge of the nuclear spin polarization can
then be used to estimate the resident electron spin
polarization. In this section, we want to estimate the
order of magnitude of the nuclear spin polarization that
can build-up in a II-VI QD and its influence on the spin
dynamics of a confined electron.

The dominant contribution to the coupling between the
confined electron and the nuclear spins originates from a
Fermi contact hyperfine interaction. This interaction can
be written as \cite{Paget1977}:

\begin{equation}
\label{Hhf} H_{hf}=\nu_0\sum_i
A^I_i|\psi(R_i)|^2(I^i_z\sigma_z+\frac{I^i_+\sigma_-+I^i_-\sigma_+}{2})
\end{equation}

\noindent where R$_i$ is the position of the nuclei $i$
with spin I$^i$ and hyperfine interaction constant A$^I_i$.
$\sigma$ and I$^i$ are the spin operators of the electron
and nuclei respectively. $\nu_0$ is the volume of the
unitary cell containing Z=2 nuclei (one Cd and one Te).
This Hamiltonian can be decomposed in a static part
affecting the energy of the electron and nuclear spins and
a dynamical part proportional to
$(I^i_+\sigma_-+I^i_-\sigma_+)$, allowing for the transfer
of angular momentum between the electron and nuclear spin
system. The static part of the hyperfine interaction leads
to the notion of effective magnetic field, either seen by
the electron due to the spin polarized nuclei (Overhauser
field $\overrightarrow{B}_N$), or by a nucleus at position
R$_i$ due to a spin polarized electron (Knight field
$\overrightarrow{B}_e^i$). These fields are defined by the
electron-nuclei interaction energy:

\begin{equation}
\label{Hhf}
H_{hf}=g_e\mu_B\overrightarrow{\sigma}.\overrightarrow{B}_N=-\sum_i\mu_I^i\overrightarrow{I}^i.\overrightarrow{B}_e^i
\end{equation}

\noindent where $g_e$ is the Lande factor of the electron
and $\mu_I^i$ the magneton of nucleus $i$ with spin $I^i$
defined by $\mu_I^i=\hbar\gamma_I^iI$ with $\gamma_I^i$ the
gyromagnetic ratio of the nucleus $i$.

\subsection{Overhauser field in a CdTe/ZnTe quantum dot}

The maximum Overhauser field resulting from a complete
polarization of the nuclei, $B_N^{max}$, is defined by
intrinsic parameters characterizing the material and the
hyperfine interaction inside the material
\cite{Testelin2008,Maletinsky2007}. The Overhauser field
can be written as:

\begin{equation}
\label{BN}
B_N=\frac{\nu_0}{g_e\mu_B}\sum_i^{N_I}A^I_i|\psi(R_i)|^2\langle I_z^i\rangle
\end{equation}

\noindent where the sum runs over N$_I$, the number of
nuclei carrying a spin I. When all the Cd and Te nuclear spins
are polarized and if we assume a homogeneous
electron wave function $\psi(R)=\sqrt{2/(\nu_0N_L)}$, with
N$_L$ the total number of nuclei in the QD, the nuclear
field reads:

\begin{equation}
\label{BNmax}
B_N^{max}=\frac{1}{g_e\mu_B}(I^{Cd}A^{Cd}p^{Cd}+I^{Te}A^{Te}p^{Te})
\end{equation}

\noindent The nuclear spin of Cd and Te are
$I^{Cd}$=$I^{Te}$=1/2 and their corresponding abundance
$p^{I}=N_I/N_L$ are $p^{Cd}=0.25$ and $p^{Te}=0.08$ (see
table \ref{table1}). Taking the hyperfine coupling
constants $A^{Cd}\approx$-31$\mu$eV,
$A^{Te}\approx$-45$\mu$eV from reference 5 and an average
electron Lande factor $g_e\approx-0.5$
\cite{Besombes2000,Leger2007}, we obtain
$B_N^{max}\approx$200mT.

\begin{table}[htb] \centering
\label{tableau1}\renewcommand{\arraystretch}{1.5}
\begin{tabular}{p{1.5cm}|p{2.5cm}|p{2.0cm}|p{2.0cm}}\hline\hline
&  Abundance ($\%$) & $I$ & $\mu_I$\\
\hline\hline $^{111}$Cd & $12.75$ & $1/2$ & $-0.5943$ \\
$^{113}$Cd & $12.26$ & $1/2$ & $-0.6217$ \\ \hline
$^{123}$Te & $0.87$ & $1/2$ & $-0.7357$ \\   $^{125}$Te &
$6.99$ & $1/2$ & $-0.8871$\\ \hline
\end{tabular}
\caption{Isotopic abundance, nuclear spin $I$ and magneton
of the nucleus $\mu_I$ for Cd and Te alloys \cite{AIP1972}.
$\mu_I$ is given in unit of the nuclear magneton $\mu_N$.}
\label{table1}
\end{table}

The Overhauser field really obtained in a QD under optical
pumping, $B_N$, is proportional to the average nuclear spin
polarization $\langle I_z\rangle$ and reaches $B_N^{max}$
when $\langle I_z\rangle$=1/2. As the electron Lande factor
in CdTe/ZnTe QDs is negative and the hyperfine constants
are negative, the sign of $B_N$ is fixed by the sign of
$\langle I_z\rangle$ which is given by the average electron
spin polarization \cite{OpticalOrientation,Testelin2008}
$\langle S_z \rangle$ along the QD growth axis $z$
($\langle S_z\rangle$=1/2 for fully polarized spin up
electrons). In the present study, the resident electron is
pumped down ($\langle S_z\rangle\leq$0) for a $\sigma$+
excitation (spin $|\downarrow\rangle$ electron) in the
presence of a positive external magnetic field. Thus, a
$\sigma$+ excitation leads to an Overhauser field, $B_N$,
antiparallel to the applied magnetic field ($B_N<0$).

\subsection{Knight field in a CdTe/ZnTe quantum dot}

At zero external magnetic field, the formation of a nuclear
spin polarization is only possible if the effective field
induced by the electron spin on the nuclei exceeds the
local field B$_l$ created by the nuclear dipole-dipole
interaction \cite{OpticalOrientation}. The Knight field is
inhomogeneous across the nuclear ensemble because the
electron wave function is not constant across the QD. In
the core of the QD where the Knight field is the strongest,
the spin diffusion induced by the nuclear dipole-dipole
coupling is suppressed and it is there that the nuclear
spins may become polarized. The magnitude of the time
averaged Knight field for a nucleus with a hyperfine
constant A$^I$ at the position R$_i$ is given by

\begin{equation}
\label{Be}
B_e^i=-\nu_0\frac{A^I}{\mu_I}|\psi(R_i)|^2(\langle
S_z\rangle f_e)
\end{equation}

\noindent  where $f_e$ is the probability that the dot is
occupied by an electron. Considering a constant
electron/nuclei overlap (homogeneous wave function
$\psi(R)=\sqrt{2/(\nu_0N_L)}$) one can obtain
\cite{Paget1977,Makhonin2010} the maximum Knight field for
nuclei with a hyperfine constant A$^I$:

\begin{equation}
\label{Bemax} B_e^{max}=-\frac{A^I}{N_L\mu_I}
\end{equation}

With $\mu_{Cd}\approx-0.6$ and a total number of nuclei in
the QD assumed to be $N_L\approx 8 \times 10^3$,
$B_e^{max}\approx$ 100 mT is derived for Cd nuclei. A
Knight field of 10mT is typical for InAs/GaAs QDs
\cite{Lai2006}. The difference is a consequence of the
smaller QD size in II-VI materials. However, B$_e$ follows
the distribution of the electron wave function in the dot
leading to a nuclear site-dependent field varying across
the dot and in optical pumping experiments, only a
weighted-averaged value of the Knight field can be
accessible.

\subsection{Dynamic nuclear spin polarization}

Under circularly polarized CW excitation, the rate equation
describing the nuclear spin polarization $I_z$ in a singly
negatively charged QD can be written as \cite{Abragam}:

\begin{eqnarray}
\label{I} \frac{\partial I_z}{\partial t}=
\frac{1}{T_{1e}}(\frac{4}{3}I(I+1)S_0-I_z)-\frac{1}{T_{dd}}I_z-\frac{1}{T_{r}}I_z
\end{eqnarray}

\noindent The last term on the right side of the equation
accounts for any spin relaxation mechanism except the
dipole-dipole interaction between nuclei which is described
by the relaxation time $T_{dd}$. The first term corresponds
to the transfer of angular momentum between the spin of the
electron and the nuclear spin bath, with $S_0$ the
polarization rate of the injected electrons and T$_{1e}$
the probability of an electron/nuclei spin "flip-flop"
given by \cite{Tartakovskii2007}:

\begin{equation}
\frac{1}{T_{1e}}=\frac{1}{T_{1e}^0}\frac{1}{1+(\Delta
E_{eZ}/\hbar)^2\tau_e^2}
\end{equation}

\noindent Here $\tau_e$ is the correlation time of the
electron/nuclear spin interaction \cite{Abragam}. $\Delta
E_{eZ}=g_e\mu_B(B_{ext}+B_N)$ is the electron Zeeman
splitting which depends on the external magnetic field,
B$_{ext}$, and the effective nuclear field B$_N$. This term
provides a feedback process mechanism between the spin
transfer rate and the nuclear polarization leading to the
enhancement of flip-flop processes when B$_N$ reduces the
electron Zeeman splitting. This feedback is responsible for
the bi-stability in the nuclear spin polarization observed
under magnetic field in InAs/GaAs QDs
\cite{Eble2006,Tartakovskii2007,Maletinsky2007}.
$T_{1e}^0$, is given by:

\begin{equation}
\frac{1}{T_{1e}^0}=f_e\tau_e(\frac{E_{hf}}{\hbar})^2
\end{equation}

\noindent with

\begin{equation}
E_{hf}=\nu_0A_i^I|\psi(R_i)|^2
\end{equation}

\noindent the interaction energy between the electron and
nuclear spin I at position $R_i$. $T_{1e}^0$ corresponds to
the nuclear spin relaxation induced by the electron at zero
electron splitting. For a homogeneous electron wave
function ($\psi(R)=\sqrt{2/(\nu_0N_L)}$), we obtain
$1/T_{1e}^0=f_e\tau_e(2A^I_i/(N_L\hbar))^2$.

The contribution of the dipole-dipole interaction to the
relaxation process is given, in the presence of an external
magnetic field, B$_{ext}$, by:

\begin{equation}
\label{pr}
\frac{1}{T_{dd}}=\frac{1}{T_{dd}^0}\frac{B_l^2}{(B_{ext}+B_e)^2+B_l^2}
\end{equation}

\noindent where B$_l$ is the local field describing the
nuclear spin-spin interaction and T$_{dd}^0$ the
characteristic time of this interaction at zero field
\cite{Dyakonov2008,Cherbunin2009,Gammon2001}. This formula
describes the acceleration of the nuclear spin relaxation
when an applied magnetic field compensates the Knight
field.

\begin{figure}[hbt]
\includegraphics[width=3.0in]{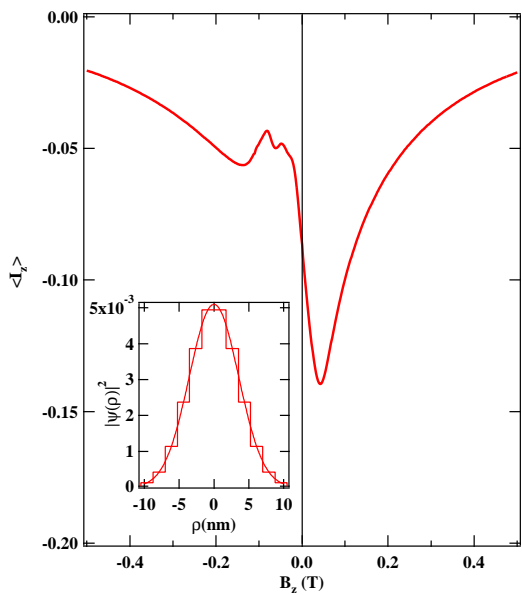}
\caption{Nuclear spin polarization obtained from equation
(\ref{I}) for a single family of nuclei with an average
hyperfine coupling A$_{av}$=-40$\mu$eV and abundance
p$_{av}$=16\% and with the parameters: f$_e$=0.4,
$S_0$=-0.4, T$_{dd}^0$=5$\mu$s, B$_l$=2.5mT, $d$=2.5nm,
$\xi$=5nm, $\tau_e$=5ns, T$_{r}$=5ms. The inset shows the
approximation of the Gaussian electron density with
$\xi$=5nm and L$_z$=2.5nm by a step function used in the
numerical calculation.}\label{fig0}
\end{figure}

The magnetic field dependence of the nuclear spin
polarization for a single family of nuclei with an average
hyperfine coupling A$_{av}$=-40$\mu$eV and abundance
p$_{av}$=16\% obtained by a numerical determination of the
steady state of the rate equation (\ref{I}) is presented in
Fig.~\ref{fig0}. In this model, the electron wave function
is described by a Gaussian function in the QD plane and a
constant function in the $z$ direction:

\begin{equation}
\label{psi}
\psi(\rho,z)=\frac{1}{\sqrt{L_z}}\frac{1}{\xi\sqrt{\pi}}e^{-\frac{\rho^2}{2\xi^2}}
\end{equation}

\noindent where $L_z$ is the thickness of the QD and $\xi$ the lateral extension of the electron wave
function. For the numerical calculation, the Gaussian wave
function is approximated by a 7 steps function (inset of
Fig.\ref{fig0}) and the 7 coupled differential equations
corresponding to the 7 families of nuclear spins ({\it
i.e.} with different Knight fields) are solved
simultaneously.

The acceleration of the dipole-dipole interaction between
the nuclear spins when the external field compensate the
Knight field is responsible for the decrease of the nuclear
spin polarisation observed at low negative magnetic field
(between $B_z\approx-50mT$ and $B_z\approx-100mT$ ). The
position of this minimum depends on the average electron
spin polarization and is a measurement of the mean value of
the Knight field. The nuclear spin polarization does not
drop to zero because of the inhomogeneity of the Knight
field: the condition B$_{tot}=0$ is satisfied only for a
small number of nuclei at any given B$_{ext}$. Provided
B$_e$ $\gg$B$_l$, the majority of nuclei experience
negligible change in depolarization.

The feedback process occurring when the Overhauser field
compensates the applied magnetic field leads to a strong
increase of the nuclear spin polarization at low positive
magnetic field ($B_z\approx50mT$). This resonant effect is
enhanced by a long electron correlation time $\tau_e$, {\it
i.e.} a weak broadening of the electron transition. This
time of free coherent electron-nuclei precession is likely
to be controlled in our experimental condition ({\it i.e.}
chemically doped QDs under CW excitation) by the
non-resonant optical injection of an electron-hole pair: we
chose $\tau_e=5ns$ in the calculation presented in
Fig.\ref{fig0}.

The important variations of the nuclear spin
polarization observed in Fig.~\ref{fig0} for a small
varying magnetic field around B$_z$=0T are expected to
significantly influence the spin dynamics of the resident
electron. In particular, an increase of the relaxation rate
of the electron spin should be observed when the external
magnetic field compensates the Overhauser field increasing
the influence of the fluctuating nuclear field.

\section{Spin injection in negatively charged quantum dots}

\subsection{Polarized fine structure of the excited state of the charged exciton}

\begin{figure}[hbt]
\includegraphics[width=3.4in]{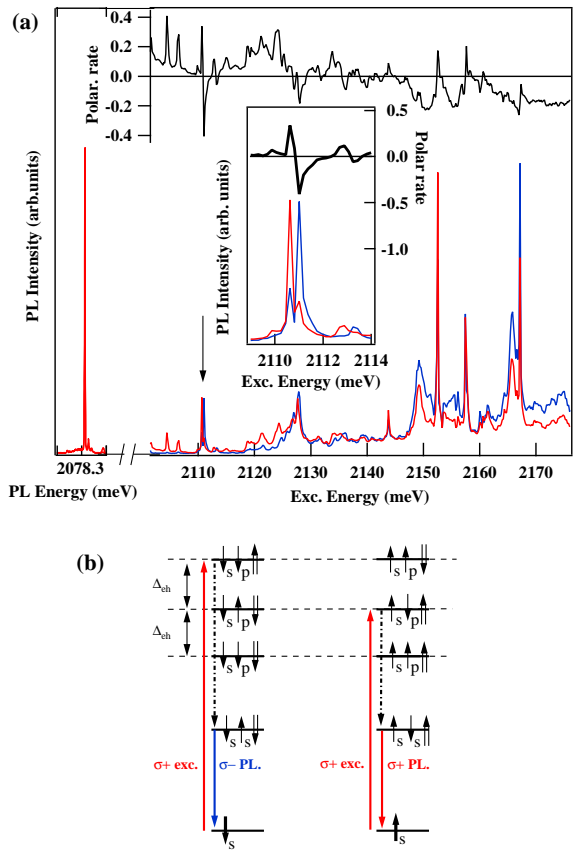}
\caption{ (a) PL and PLE spectra resolved in circular
polarisation under $\sigma+$ CW excitation. The inset is a
zoom on the polarized doublet in the PLE spectra, the
polarisation rate is also displayed. (b) Energy levels of
the negative trion states. The electrons triplet state is
split by the electron-hole exchange energy $\Delta_{eh}$.
The electrons singlet state, also part of the p-shell is
not represented here. Left scheme: when exciting with
$\sigma +$ light on the triplet state
$\left|S=1,S_z=-1\right\rangle$, photon absorption occurs
only if the resident electron is down. As demonstrated by
M.E. Ware \cite{Ware2005}, during the excited trion
relaxation, an electron-hole flip-flop process allowed by
anisotropic exchange interactions results in a $\sigma -$
PL. Right scheme: when exciting with $\sigma +$ light on
the triplet state $\left|S=1,S_z=0\right\rangle$, photon
absorption occurs only if the resident electron is up. Fast
relaxation from this state leads to $\sigma +$
PL.}\label{fig1}
\end{figure}

In order to prepare the spin state of a resident carrier in a singly charged QD, spin polarized
electron-hole pairs are injected through circularly
polarized photo-excitation of an excited state of the QD. Low power PLE
spectra on a singly negatively charged QD presented in
Fig.~\ref{fig1} reveals intense absorption resonances for
X$^-$ with a strong polarization dependence. In general we
find three distinctive features in these excitation
spectra.

The first is a set of lower energy resonances that are
strongly co-polarized with the excitation laser. These
transitions can be assigned to nominally forbidden
transitions involving states with two s-shell electrons and
an excited or delocalized hole. These transitions are
particularly well observed in CdTe/ZnTe structures because
of the weak valence band offset.

The second feature is a higher energy resonance that
displays a fine structure doublet well resolved in circular
polarization. As presented in Fig.~\ref{fig1}(a), the PLE
exhibits a strongly co- and then cross-polarized resonance
as the laser  energy increases around 2110 meV. As proposed
by M.E. Ware {\it et al.} \cite{Ware2005} we can assign
this doublet to the direct excitation of the two bright
triplet states (see Fig.~\ref{fig1}(b)) of the excited
negatively charged exciton ($X^{-*}$). $X^{-*}$ consists of
an electron-hole pair in the $S$-shell and an electron in
the $P$-shell. This doublet is a characteristic signature
of the presence of a single electron in the QD ground
state. We have found a triplet splitting $\Delta_{eh}$
around 400 $\mu$eV changing from dot to dot. This is higher
than the values found in InAs QDs, in agreement with the
stronger exchange interaction in our II-VI QD system.

For an excitation above the $X^{-*}$ triplet states, a
series of excited states and an absorption background with
a significant negative circular polarisation rate are
observed. As we will discuss in the next section, this
condition of excitation can be used to perform an optical
pumping of the resident electron spin.

\subsection{Kinetics of the degree of circular polarization}

As a probe of the resident electron spin orientation, we
will use the amplitude of the negative circular
polarization of the charged QDs \cite{Ikezawa2005,
Braker2005,Shabaev2009}. In the case of $X^-$, the circular
polarization of the emitted light reflects both the spin of
the resident electron before the absorption of a photon and
the spin of the hole before emission. Negative polarization
of X$^-$ implies that the hole spin has flipped prior to
recombination and that a spin flipped hole contributes to
the X$^-$ formation with a higher probability than a
non-flipped hole. This process also leads to an optical
pumping of the resident electron spin.

  \begin{figure}[hbt]
\includegraphics[width=3.5in]{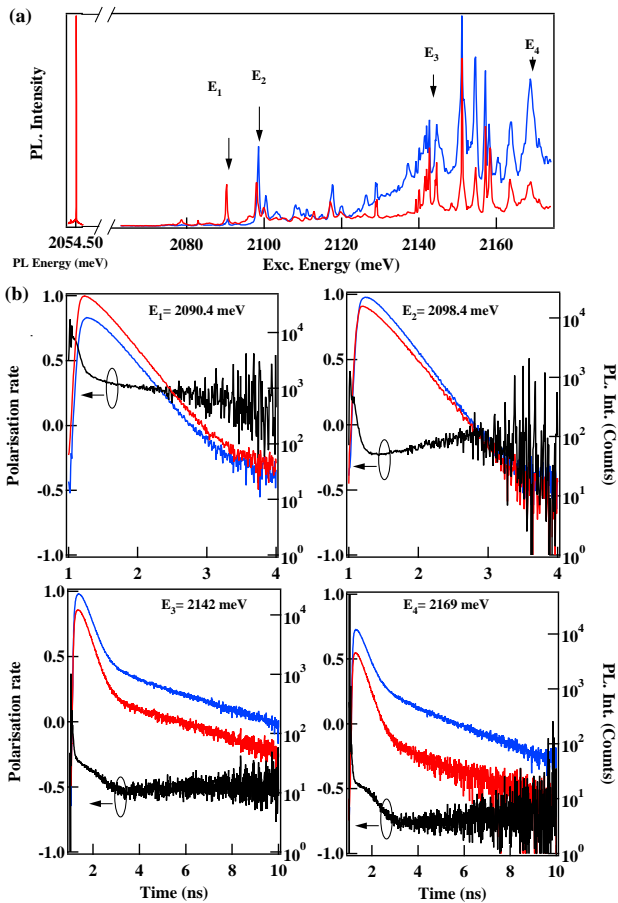}
\caption{(a) PL and PLE spectra of a singly charged quantum
dot. (b) Time-evolution of the co-polarized PL (red line),
the cross-polarized (blue line) PL, and the degree of
circular polarisation (black line) under pulsed excitation
for four different excitation energies ($E_{1}$ to
$E_{4}$). A clear negative polarization rate is
observed when the excitation energy is higher than the
triplet states of the excited charged exciton X$^{-*}$
($E_{2}$).}\label{fig2}
\end{figure}

The kinetics of the degree of circular polarisation
observed in time resolved PL experiments reflects the
mechanisms of spin injection. The time dependence of the
polarisation rate $\rho_c$ under quasi-resonant pulsed
excitation ($\approx$ 2ps) is displayed in Fig.~\ref{fig2}.
As we discussed for the PLE spectra of the QD presented in
Fig.~\ref{fig1}, the lowest excited state (E$_{1}$ in
Fig.~\ref{fig2}) is fully co-polarized. It corresponds to
the negatively charged state with two electrons in the
$S$-shell and a hole in a higher shell. It is spin
selective as the spin of the two electrons in the $S$-shell
have to be opposite: under $\sigma+$ excitation, an
absorption only occurs if the resident electron is
$|\uparrow\rangle$. The fast initial decay (in the range of
100 ps) of the polarization under excitation on E$_{1}$ is
attributed to a relaxation of the hole spin on the high
energy shell while the slower decay (in the range of 5 ns)
is attributed to relaxation of the hole spin when it is in
the $S$-shell.

Resonant $\sigma+$ excitation on the charged exciton
triplet states (E$_{2}$ in Fig.~\ref{fig2}) results in the
creation of an excited trion
$\Uparrow_P\downarrow_P\uparrow_S$ or
$\Uparrow_P\downarrow_P\downarrow_S$ depending on the spin
of the resident electron (see Fig.~\ref{fig1}(b)). Under
excitation on E$_{2}$ we observe a fast decay of the
initial positive polarization rate: the polarization rate
becomes negative in a few tens of $ps$. This evolution
reflects the spin dynamics of $X^{-*}$. The excited state
$\Uparrow_P\downarrow_P\uparrow_S$ can relax quickly to the
ground trion state while for the state
$\Uparrow_P\downarrow_P\downarrow_S$, relaxation to the
ground state is forbidden until an electron-hole flip-flop
occurs through anisotropic exchange interaction
\cite{Cortez2002}. Therefore, relaxation from $X^{-*}$
results in a positive polarization rate at short delays and
negative at longer delays.

For higher excitation energies (E$_{3}$ and E$_{4}$ in Fig.
\ref{fig3}) a major part of the decay of the polarization
rate takes place within the first 200 ps. The polarization
rate becomes quickly negative and approches a value of
about -30$\%$. It further decreases at longer time delay
($\approx 1ns$) and reaches a steady state value lower than
-50$\%$. To understand the three regimes in the dynamics of
the polarization rate, we have to consider 0D-2D
cross-transitions where the electron is injected in the dot
and the hole in the wetting layer. Such transitions are
particularly important in our system presenting a weak
valence band offset. The spin of the hole is randomize
before its capture by the QD whereas the electron spin is
conserved (spin $|\downarrow\rangle$ for $\sigma+$
excitation). The captured bright excitons ({\it i.e.}
without hole spin flip) or dark excitons ({\it i.e.} after
a hole spin flip) relax to form the hot trion $X^{-*}$. For
an unpolarized resident electron, four possible channels
are then possible for the relaxation of the hot trion
formed by a $\sigma+$ excitation:

\begin{enumerate}
\item $\uparrow_s\downarrow_p\Uparrow$ $\longmapsto$ $\uparrow_s\downarrow_s\Uparrow$ $\longmapsto$ $\sigma+$ $and$ $\uparrow_s$
\item $\uparrow_s\downarrow_p\Downarrow$ $\longmapsto$ $\uparrow_s\downarrow_s\Downarrow$ $\longmapsto$ $\sigma-$ $and$ $\downarrow_s$
\item $\downarrow_s\downarrow_p\Uparrow$ $\stackrel{\delta_a}{\longmapsto}$ $\downarrow_s\uparrow_s\Downarrow$ $\longmapsto$ $\sigma-$ $and$ $\downarrow_s$
\item $\downarrow_s\downarrow_p\Downarrow$ $\stackrel{\tau_h}{\longmapsto}$ $\downarrow_s\downarrow_p\Uparrow$ $\stackrel{\delta_a}{\longmapsto}$ $\uparrow_s\downarrow_s\Downarrow$ $\longmapsto$ $\sigma-$ $and$ $\downarrow_s$
\end{enumerate}

\noindent where $\delta_a$ is an anisotropic electron-hole
exchange interaction term responsible for the flip-flop of
the electron-hole pair in the QD excited state and $\tau_h$
the spin flip time of a hole in the $S$-shell. The
realization of (1) and (2) is proportional to the
probability for the resident electron to be
$|\uparrow\rangle$ and does not require any spin flip of
the hot trion: they take place in the $ps$ range. (3) and
(4) depend on the probability of having the resident
electron $|\downarrow\rangle$ and involve spin flips of the
hot trion. These last two channels ((3) and (4)) lead to
the appearance of a negative circular polarization rate
with two time scales, one in the tens of $ps$ range
governed by $\delta_a$ and one in the $ns$ range governed
by $\tau_h$.

If the spin relaxation rate of the resident electron is
longer than the optical excitation rate, cumulative effects
lead to the optical orientation of the electron spin. At
this stage, we have to notice that the observation of
negative circular polarization does not necessarily mean
that the resident electron is polarized but a variation of
its polarization will cause a change in the negative
circular polarization rate.

\section{Electron spin optical orientation}

\subsection{Orientation of the spin of the electron}

\begin{figure}[hbt]
\includegraphics[width=3.5in]{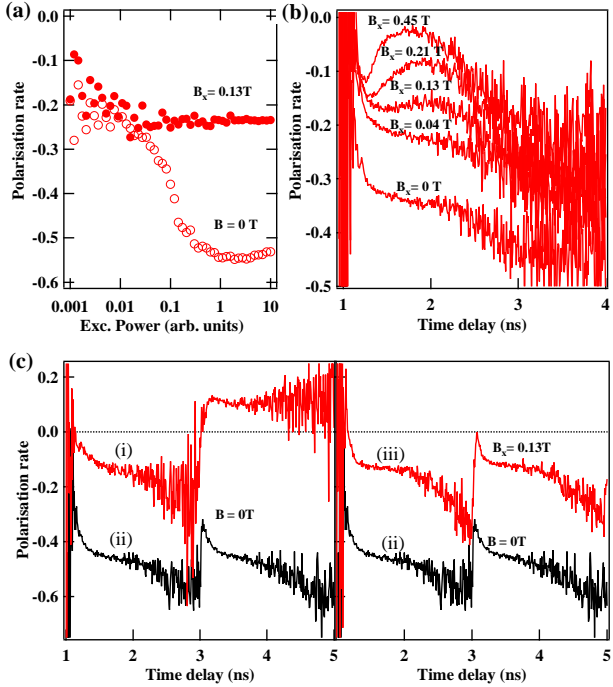}
\caption{(a) Evolution of the degree of circular
polarisation as a function of CW excitation power. Empty
circles are zero field measurements, filled circles are
measurements in a transverse magnetic field B$_x$=0.13T.
(b) Time-evolution of the degree of circular polarisation
under pulsed (2ps) quasi-resonant excitation for different
transverse magnetic fields. (c) Time-evolution of the
degree of circular polarisation with a two-pulse-excitation
sequence: the pulses are separated by 2ns. (i) is obtained with cross-polarized pulses at
zero field. (ii) is obtained with co-polarized pulses at
zero field. (iii) is obtained with co-polarized pulses and
a transverse magnetic field B$_x$=0.13T. }\label{fig3}
\end{figure}

The presence of optical pumping of the resident electron is
confirmed by the power dependence of the negative circular
polarization rate obtained under CW excitation
(Fig.~\ref{fig3}(a)). As the pump power intensity is
increased, we observe a rapid growth of the negative
circular polarization with a saturation at about -55$\%$.
This reveals the progressive orientation of the resident
electron spin by the exciting beam. The remaining
polarization rate observed at low excitation power or in a
weak transverse magnetic field is attributed to the different
processes of carrier relaxation discussed in section IV. In
the optical pumping regime ({\it
i.e.} without transverse magnetic field and at large
excitation intensity), the measurement of the negative
circular polarization gives an estimate of the degree of
the electron spin polarization.

The electron spin memory can be significantly erased by a
weak magnetic field (see Fig.~\ref{fig3}(b)) applied in the
plane of the QD. At B$_x$$\approx0.1T$, all the
contribution of the electron spin polarization to the
negative circular polarization rate has disappeared. Despite
the weak transverse component of the hole g-factor, a
further increase of the transverse magnetic field can
induce a precession of the confined hole spin during the
lifetime of the negatively charged exciton. At high field,
this precession depolarizes the hole spin and finishes to
destroy the average negative circular polarisation of the
$X^-$. This effect is observed in Fig.~\ref{fig3}(b) as an
oscillation of the polarization rate for a transverse field
larger than 0.1T. This oscillation corresponds to the first
period of precession of a spin polarized hole injected at
$t=0ns$. The decrease of the polarization rate at long time
delay corresponds to the late recombination of spin flipped
holes stored as dark excitons in the triplet state of
X$^{-*}$.

The dynamics of the pumping and relaxation of the resident
electron spin can be estimated through the time evolution
of the polarization rate when the QD is excited by a
sequence of two circularly polarized picosecond laser
pulses. The results of the experiments using equal
intensities for the two pulses are shown in
Fig.~\ref{fig3}(c). When the QD is excited with circularly
co-polarized pulses (ii), a large average negative circular
polarization is observed for both PL pulses. However, in
the case of excitation by cross-polarized beams (i), the
average PL polarization vanishes. These results directly
demonstrate that the spin orientation created by the first
pulse affects the polarization of the PL excited by the
second one. It means that after recombination of the
electron-hole pair, the information about the polarization
of the excitation is stored in the orientation of the
resident electron spin. In addition, we notice that the
polarization rate is identical for the two pulses in the
excitation sequence. As these pulses are separated in time
by either $2ns$ or $11ns$ this suggest that the relaxation
time of the electron exceeds by far the laser pulses
repetition rate ($\approx 13ns$). Consequently, the
resident electron can be fully depolarized by a weak
transverse magnetic field. A significant decrease of the
negative polarisation rate is observed for both pulses in a
transverse field B$_x$=0.13T (iii) confirming the influence
of the optical pumping of the electron spin on the negative
polarization rate.

\subsection{Dynamics of the electron spin orientation}

\begin{figure}[hbt]
\includegraphics[width=3.5in]{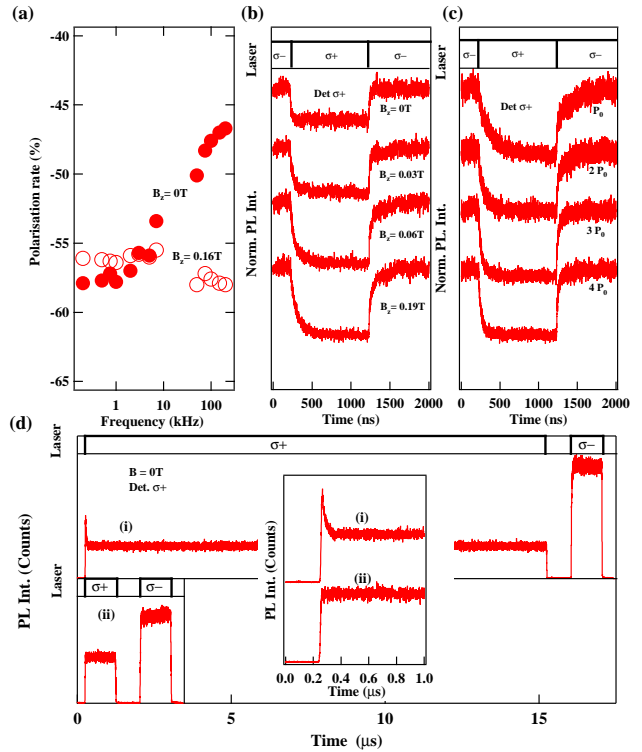}
\caption{ (a) Evolution of the NCP with the frequency of
the $\sigma+$/$\sigma-$ modulation of the excitation laser
at zero magnetic field and at a field B$_z$=0.16T applied
along the QD growth axis $z$. Time-evolution of the
$\sigma+$ PL excited alternatively with $\sigma+$ or
$\sigma-$ light for different magnetic fields applied in
Faraday geometry (b) and for different excitation
intensities and a fixed magnetic field B$_z$=0.19T (c). (d)
Time-evolution of the $\sigma+$ PL excited alternatively
with $\sigma+$ and $\sigma-$ light trains. The excitation
sequence are displayed above the spectrum. The inset is a
zoom on the fast transient at short time
delay.}\label{fig4}
\end{figure}

In the presence of optical pumping, the degree of negative
circular polarization of $X^-$ reflects the spin
polarisation of the resident electron. The dynamics of its
optical orientation can then be revealed by the observation
of the negative polarization rate under modulated
circularly polarized excitation. As presented in
Fig.~\ref{fig4}(a), the negative polarisation strongly
depends on the modulation frequency: an increase of the
degree of circular polarisation when the modulation
frequency is decreased is observed at B=0T. This modulation
frequency dependence is canceled by a magnetic field of
B$_z$=0.16T applied along the QD growth axis. As already
observed in InAs QDs, this behavior is a fingerprint of the
coupling of the electron to a fluctuating nuclear field\cite{Moskalenko2009}.

The QD contains a finite number $N_I$ of nuclei carrying a
spin, which means that statistically, the number of spins
parallel and antiparallel in any given direction differs by
a value $\sqrt{N_I/3}$. The result is an effective magnetic
field B$_f$, oriented in a random direction. This field
will induce a precession of the spin of the electron for
every B$_{f}$ not aligned along the QD growth axis $z$.
B$_f$ can be estimated, for a CdTe QD. Assuming a
homogeneous envelope-function for the electron
($\psi(R)=\sqrt{2/(\nu_0N_L)}$), B$_{f}$ is given by
\cite{Dyakonov2008}:

\begin{eqnarray}
B_f^2=\frac{2}{(g_e\mu_B\sqrt{N_L})^2}(I_{Cd}(I_{Cd}+1)A_{Cd}^2
p_{Cd}\\ \nonumber +I_{Te}(I_{Te}+1)A_{Te}^2 p_{Te})
\end{eqnarray}

For our estimation of $N_L=8000$ one obtain
$B_f\approx12mT$. The electron spin precession frequency in
the frozen nuclear spin fluctuation B$_f$ $\approx$12mT is
$\approx$80MHz. This frequency can be smaller than the rate
of optical injection of the spin polarized carriers at high
excitation intensity allowing an optical pumping of the
electron spin even in the presence of nuclear spin
fluctuations.

At high modulation frequency of the polarisation and low
excitation intensity, a dynamic nuclear spin polarisation
does not have time to build-up. Over time scales less than
1 $\mu s$, the electron is exposed to a snapshot of B$_f$
where the nuclear spin configuration remains frozen. In the
absence of an external magnetic field, only this internal
field B=B$_f$ acts on the electron. For a randomly oriented
nuclear spin system, the electron spin polarization quickly
decays to 1/3 of its initial value due to the frozen
nuclear field \cite{Merkulov2002}. This decay is not a real
relaxation process as the electron coherently evolves in a
frozen nuclear spin configuration. On an averaged
measurement, a fast decay of the electron polarization on a
characteristic timescale $t\approx h/(g_e\mu_BB_f)$ is
expected\cite{Merkulov2002}. In the absence of nuclear spin
polarization, the influence of the fluctuating nuclear
field can be suppressed by applying an external magnetic
field. For sufficiently large external fields, the nuclear
spin fluctuations does not contribute significantly to the
total field, B$_{tot}$=B$_{ext}$+B$_f$, and the
electron-spin polarization is preserved.

At small modulation frequencies of the polarization or
under CW excitation, nuclei can be dynamically oriented
through flip-flop with the spin polarized resident
electron. This nuclei orientation leads to the formation of
an Overhauser field, B$_N$ along the $z$ axis, which may be
much larger than the in-plane component of the fluctuating
field (B$_f$). The electron now precesses around a nuclear
field whose $z$ component dominates: the result is an
increase of the average electron spin polarization compared
to the case of a totally randomly oriented nuclear spin
system. The effect of the Overhauser field is similar to an
applied magnetic field along the $z$ axis allowing an
optical orientation of the electron spin even at low
excitation power.

This influence of the nuclear spin fluctuations in the
optical pumping of the electron is confirmed by the
magnetic field dependence of the time resolved polarisation
rate obtained at high modulation frequency
(Fig.~\ref{fig4}(b)). The increase of the polarization rate
and the appearance of a transient with an applied external
magnetic field along $z$ reflects an increase of the
optical pumping efficiency of the electron \cite{Feng2007}.
This optical pumping, which takes place in a few tens of
$ns$, is promoted by the presence of the external field
which can dominate the fluctuations of the Overhauser field
\cite{Petrov2008}. This short timescale component in the
dynamics of the polarization of the spin of the electron
becomes faster with the increase of the optical generation
rate of spin polarized carriers and reach the $ns$ range
(Fig.~\ref{fig4}(c)).

Similarly to the application of an external magnetic field,
the build-up of a DNSP favors the electron spin
polarization. This is confirmed by the following
experiment: in Fig.~\ref{fig4}(d), the $\sigma+$ PL has
been time-resolved using the two different excitation
sequences displayed on each spectrum. In the sequence (ii),
the excitation pulses are of equal length and power, and
are short enough to prevent the creation of DNSP. In
sequence (i), the difference of pulses length allows the
creation of DNSP. The measurements show two striking
differences. First, the average circular polarisation,
given by the difference of the PL intensity obtained under
$\sigma-$ and $\sigma+$ excitation, is higher in (i) than
in (ii). Second, the PL of (i) exhibits a fast PL transient
at short delay reflecting an optical pumping of the
electron spin (detail of this transient is shown in the
inset of Fig.~\ref{fig4}(d)). These two features
demonstrate that the Overhauser field created in (i) is
strong enough to block the longitudinal decay of the electron spin
by the fluctuating nuclear field.

\section{Nuclear spin polarisation}

\subsection{Built-up of the nuclear spin polarization}

\begin{figure}[hbt]
\includegraphics[width=3.5in]{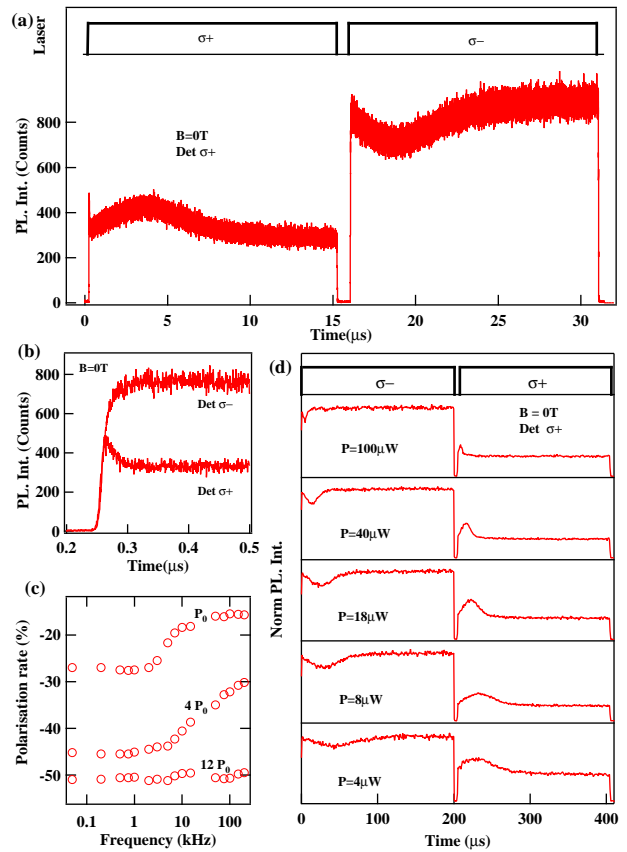}
\caption{(a) Time-evolution of the $\sigma+$ PL excited
alternatively with $\sigma+$/$\sigma-$ light trains. The
excitation sequence is displayed above the spectrum. (b)
Zoom on the transient corresponding to the optical pumping
of the electron. The time-evolution of the $\sigma-$ PL
recorded in the same conditions of excitation is also
displayed.(c) Dependence of the the degree of circular
polarisation of the QD PL on the frequency of the
$\sigma+$/$\sigma-$ modulation of the light for different
excitation power. (d) Time-evolution of the $\sigma+$ PL
excited alternatively with $\sigma+$/$\sigma-$ light for
different excitation power.}\label{fig5}
\end{figure}

Direct evidence of the build-up of a DNSP can be observed
using sequences of pulses of long duration (tens of
$\mu$s). As displayed in Fig.~\ref{fig5}(a), the $\sigma+$
PL recorded under $\sigma+$ excitation presents first a
fast transient with a drop of the intensity due to the
orientation of the resident electron spin (a zoom of the
transient at short delay is presented in
Fig.~\ref{fig5}(b)). Then, a slower transient is observed:
the $\sigma+$ PL increases during a few $\mu$s
reflecting a decrease of the absolute value of the negative
polarisation ({\it i.e.} of the spin polarization of the
resident electron) before it decreases again. This
evolution has been predicted by M. Petrov \textit{et al.}
\cite{Petrov2009} and results from a destruction of the
Overhauser field created at the end of the $\sigma-$
excitation pulse, and a build-up of an Overhauser field in
the opposite direction under $\sigma+$ excitation. During
this process, the amplitude of the Overhauser field becomes
zero and the electron spin is strongly affected by the
nuclear spin fluctuations B$_f$.

As presented in Fig.~\ref{fig5}(d), the speed of the
destruction and build-up of the nuclear polarization
strongly increases with the increase of the excitation
power. Simultaneously, the average negative polarisation of
the $X^-$, given by the intensity difference of the PL
obtained under $\sigma+$ and $\sigma-$ excitation,
increases. This effect is also directly observed in the
modulation frequency dependence of the polarization rate
displayed in Fig.~\ref{fig5}(c). The modulation frequency
required to suppress the nuclear spin polarisation ({\it
i.e.} to decrease the absolute value of the polarization)
increases with the excitation intensity. At low excitation
intensity, a formation time of the nuclear spin
polarization of about $50\mu s$ can be estimated from the
modulation frequency dependence of the negative circular
polarization \cite{Moskalenko2009}. This is three order of
magnitude faster than in InAs/GaAs QDs where a pumping time
of the nuclei around $10 ms$ has been reported
\cite{Maletinsky2007}. At hight excitation power, the
pumping rate of the nuclei becomes faster than the
polarization modulation frequency and a stable negative
polarization of about -50$\%$ is obtained
(Fig.~\ref{fig5}(c)). An increase of the value of the
negative polarization with the excitation intensity is also
observed suggesting an increase of the average nuclear spin
polarization and Overhauser field with the excitation
intensity.

\subsection{Magnetic field dependence of the nuclear spin polarisation}

\begin{figure}[hbt]
\includegraphics[width=3.2in]{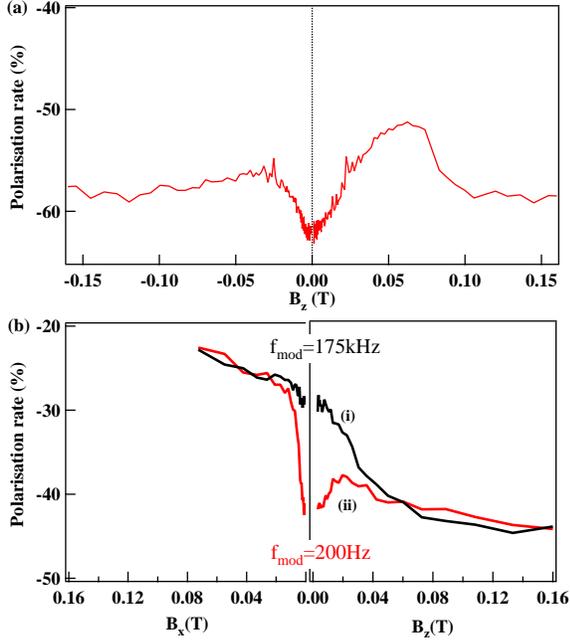}
\caption{ (a) Magnetic field dependence in Faraday
configuration of the negative polarization rate under
$\sigma+$ CW excitation. (b) Magnetic field dependence of
the negative polarization rate, on a different QD, in
Voight (left) and Faraday (right) configuration under
$\sigma+$/$\sigma-$ modulated excitation at low
(f$_{mod}$=200Hz) and high (f$_{mod}$=175kHz) modulation
frequency.}\label{fig6}
\end{figure}

A typical magnetic field dependence of the polarization
rate of a singly charged QD under CW circularly polarized
excitation, in the optical pumping regime, is presented in
Fig.~\ref{fig6}: Fig.~\ref{fig6}(a) focuses on the Faraday
geometry while the Voigt geometry and the influence of the
modulation frequency of the polarization of the excitation
beam are presented in Fig.~\ref{fig6}(b). The asymmetry of
the response in the Faraday geometry is a fingerprint of
the presence of a nuclear spin polarization induced by the
helicity of the excitation beam.

A striking feature of this magnetic field dependence is the
small increase of the absolute value of the negative
polarization around B=0T. This is in opposition to what is
usually observed in III-V semiconductor QDs where a
decrease of the electron spin polarisation occurs at weak
magnetic field because of the dominant contribution of the
fluctuating nuclear field B$_f$. As presented in curve (i)
of Fig.~\ref{fig6}(b), a standard increase of polarisation
with magnetic field is restored in the absence of nuclear
polarisation ({\it i.e.} under polarization modulated
excitation). This shows that the increase of the electron
polarization around $B=0T$ observed under CW excitation is
linked to the DNSP: the nuclear spin fluctuations are
strongly suppressed by the build-up of a large Overhauser
field.

The experiment presented in Fig.~\ref{fig6}(a) was carried
out under CW $\sigma+$ excitation, pumping the resident
electron down. This leads to an average polarization of
nuclei with $\left\langle I_z\right\rangle<0$ and an
Overhauser field $B_N\leq0$. For $B_z\geq0$, we observe
around 50mT an increase of the circular polarization rate
of $10\%$ which reflects a depolarization of the resident
electron. This behavior is attributed to the compensation
of the Overhauser field by the external Faraday field
($B_{z}=-B_N$). As the electron precesses around the total
field $B_{tot}=B_{z}+B_N+B_f$, the electron dynamics is
then governed by the nuclear spin fluctuations $B_f$,
resulting in a depolarization of the resident electron.

Considering the left panel of Fig.~\ref{fig6}(a)
(corresponding to $B_z\leq0$), we observe at $B=0T$ a
maximum in the electron polarization, then a decrease of
about $5\%$ in the first 25 mT followed by a small increase
at larger fields. This extremum around $B_z=-25mT$ also
reflects a depolarization of the resident electron spin.
This depolarization is attributed to a compensation of the
Knight field by the external magnetic field. The nuclear
field is then close to zero and the electron dynamics is
ruled by the sum of $B_{z}\approx25mT$ and the nuclear spin
fluctuations $B_f$. The effect of $B_f$ is not negligible
at $25mT$. As observed in the experiment of
Fig.~\ref{fig4}(b), in the absence of DNSP, the
polarization of the PL continuously increases as the
Faraday magnetic field is increased from $0$ to $60mT$.

This influence of the Knight and Overhauser fields is
further-confirmed by the following experiment: We present
on the right panel of Fig.~\ref{fig6}(b) a measurement,
were we have studied the polarization rate as a function of
the magnetic field $B_z$ under modulated excitation. The
light is modulated $\sigma+/\sigma-$ at two different
rates: (i) $175$ kHz ($\approx3\mu s$ of $\sigma+$ exc.,
then $\approx3\mu s$ of $\sigma-$ exc. of equal intensity)
and (ii) $200$ Hz ($\approx2500\mu s$ for a given
polarization). Hence, at low power of excitation, DNSP is
achieved in (ii) and not in (i). The detection is done on
an APD synchronized with the modulation, and for a fixed
circularly polarized detection. We measure a polarization
rate by varying the excitation polarization and not the
detection. Hence, this polarization rate is an average of
the polarization rates measured in Fig.~\ref{fig6}(a) for
$B_z$ and $-B_z$. The magnetic field dependence for (ii) is
consistent with the one observed in the CW regime, with an
evolution ruled by the competition between the $B_z$,
$B_N$, $B_e$ and $B_f$. On the other hand the magnetic
field dependence (i) is only controlled by the competition
between $B_z$ and $B_f$. For sufficiently large external
fields, the nuclear spin fluctuations $B_f$ do not
contribute to the total field and the electron-spin
polarization does not decay. The width at half maximum is
$25\pm5mT$. This gives an order of magnitude of the
fluctuating Overhauser field.

\begin{figure}[hbt]
\includegraphics[width=3.2in]{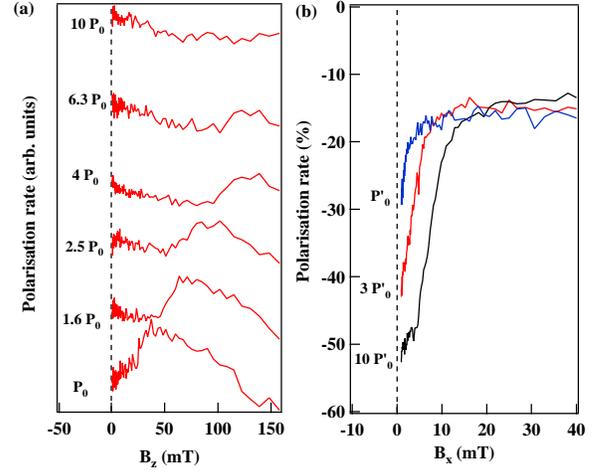}
\caption{Excitation power dependence of the negative
polarization rate under magnetic field. (a) Faraday
configuration under $\sigma$+ CW excitation. (b)  Voigt
configuration.}\label{fig7}
\end{figure}

The magnetic field dependence of the polarization rate
under CW excitation has been performed for $B_z>0$ for
different excitation powers (Fig.~\ref{fig7}(a)). With
increasing power, the minimum in the electron polarization
is shifted to higher magnetic field, evidencing an increase
of the polarization of the nuclei and of the resulting
Overhauser field. At high excitation intensity, a
significant portion of the nuclei are polarized and the
minimum of electron polarization is observed around
$B_z=-B_N=100mT$ which would correspond to $50\%$ of the
maximum Overhauser field. However,
the parameters of II-VI QDs used to estimate this maximum
field are not known with precision and this percentage is
subject to caution.

\section{Dynamics of coupled electron and nuclear spins.}

\subsection{Electron-nuclear spin system in a transverse
magnetic field}

A transverse magnetic field dependence of the polarization
rate of X$^-$ is presented in the left panel of
Fig.~\ref{fig6}(b). As the transverse magnetic field is
increased, we observe in the absence of DNSP (black curve,
corresponding to fast $\sigma+/\sigma-$ modulation), a
progressive decrease of the negative polarization rate over
the first $80mT$. For a spin polarized electron and in the
absence of nuclear spin polarization, two processes can
contribute to the observed Hanle depolarization of X$^-$.
The first is a depolarization of the resident electron
governed by the transverse relaxation time of the spin of
the electron, $T_2$, in an unpolarized nuclear spin bath
(standard Hanle depolarization). This $T_2$ should give
rise to a half width of the Hanle curve
$B_{1/2}=B_f\approx25mT$, deduced from the Faraday
measurement in Fig.~\ref{fig6}(b). The second mechanism is
a precession of the hole during the charged exciton
life-time. This process is expected to play a role above
$50mT$ as we have seen in the time resolved polarization
rates presented in Fig.~\ref{fig3}(b). It is not possible
to discriminate between the two mechanisms as, because of
the weak polarization of the electron, they are both
responsible of a small decrease of few $\%$ of the circular
polarization rate.

More interesting is the comparison with the data where a
DNSP is created (red curve on the left panel of
Fig.~\ref{fig6}(b)). In this later case, a fast decrease of
the negative polarization rate is observed when increasing
the transverse field from $0$ to $10mT$. The half width at
half maximum of the depolarization curve is $\approx5mT$.
This efficient depolarization of the resident electron is
due to the precession of the coupled electron-nuclei system
\cite{OpticalOrientation, Oulton2007, Cherbunin2009}. After
this fast depolarization of the electron, the negative
polarization rate reaches the value observed in the absence
of DNSP ({\it i.e.} under fast $\sigma+$/$\sigma-$
modulated excitation, black curve). This is also observed
in the modulation frequency dependence of the polarization
rate presented in Fig.~\ref{fig3}(a).

We observe that the depolarization curve in transverse
magnetic field in the presence of nuclear spin polarization
(Fig.~\ref{fig7}(a)) strongly depends on the excitation
power and deviates from a Lorentzian shape at high
excitation power. For the later case, the negative
polarization rate seems to be weakly affected by the first
few $mT$ of transverse magnetic field, and then decreases
abruptly.

The width of such depolarization curve is typically 50mT in
a singly charged InAs/GaAs QDs in the presence of nuclear
spin polarization \cite{Desfonds2010}. In InAs QDs, an
influence of the magnetic anisotropy of the nuclei produced
by the in-plane strain is also observed in the transverse
magnetic field dependence of the Overhauser field
\cite{Dzhioev2007, Krebs2010}. This cannot be the case in
CdTe QDs as the nuclear spins I=1/2 for Cd and Te.

\begin{figure}[hbt]
\includegraphics[width=3.5in]{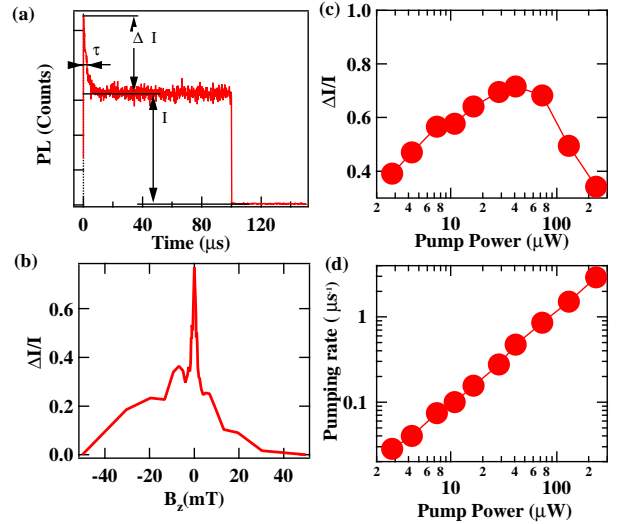}
\caption{Variation of the amplitude (c) and characteristic
time (d) of the DNSP transient under $\sigma+$ excitation
(an exemple is presented in (a)) as a function of the
excitation intensity at B$_z$=0mT with a constant dark time
$\tau_{dark}=50\mu s$. (b) Variation of the amplitude of
the DNSP transient with a magnetic field applied along the
QD growth axis.}\label{fig9}
\end{figure}

The power dependence observed in our system could arise as
the electron is pumped faster than the precession in the
transverse applied ($\tau_{press}^e=4ns$ at $B_x=5mT$). At
hight excitation intensity, the resident electron is
replaced by an injected spin polarized electron faster than
the precession in the transverse magnetic field. The
depolarization of the electron spin is then given by the
power dependent Hanle curve:

\begin{equation}
\label{Hanle}
S_{z}(\Omega)=\frac{S_z(0)}{(1+(\Omega\tau)^2)}
\end{equation}

\noindent where $\Omega=g_e\mu_BB/\hbar$ and
$1/\tau=1/\tau_p+1/\tau_s$ with $1/\tau_p$ the pumping rate
and $1/\tau_s$ the electron spin relaxation rate. However,
this power broadening does not explain the slight deviation
form the Lorentzian shape we observed at high excitation
intensity in (Fig.~\ref{fig7}(b)).

The creation of DNSP could also be faster than the nuclei
precession ($\tau_{press}^N=5\mu s$ at $B_x=5mT$). The
decrease of the electron polarization in a transverse
magnetic field can then be influenced by the decrease of
the steady state nuclear field. As a result of this
decrease, the total in-plane component of the magnetic
field which controls the electron precession increases more
slowly than the external field B$_x$: the precession of the
electron would be efficiently blocked by the Overhauser
field, and the electron polarization would be conserved.
Such a scenario would be specific to II-VI quantum dots,
where the build-up of DNSP is fast enough to block the
precession of nuclei. This point requires further
investigation.

\subsection{Dynamics of the nuclear spin polarization}

In order to analyze quantitatively the build-up time and
the characteristic amplitude of the  polarization transient
induced by the DNSP, we perform a time-resolved measurement
using a $100\mu s$ pulse of a $\sigma+$ helicity, followed
by a $50\mu s$ dark time during which the DNSP relaxes
partially (quantitative analysis of this relaxation will be
done in the next section and is indeed found to occur on a
time-scale shorter than $50\mu s$ for $B_z<5mT$). This
experimental configuration enables us to fit the observed
DNSP transient by an exponential variation, permitting to
extract a characteristic rate $1/\tau$ and amplitude
$\Delta I$ (Fig.~\ref{fig9}(a)).

\begin{figure}[hbt]
\includegraphics[width=3.5in]{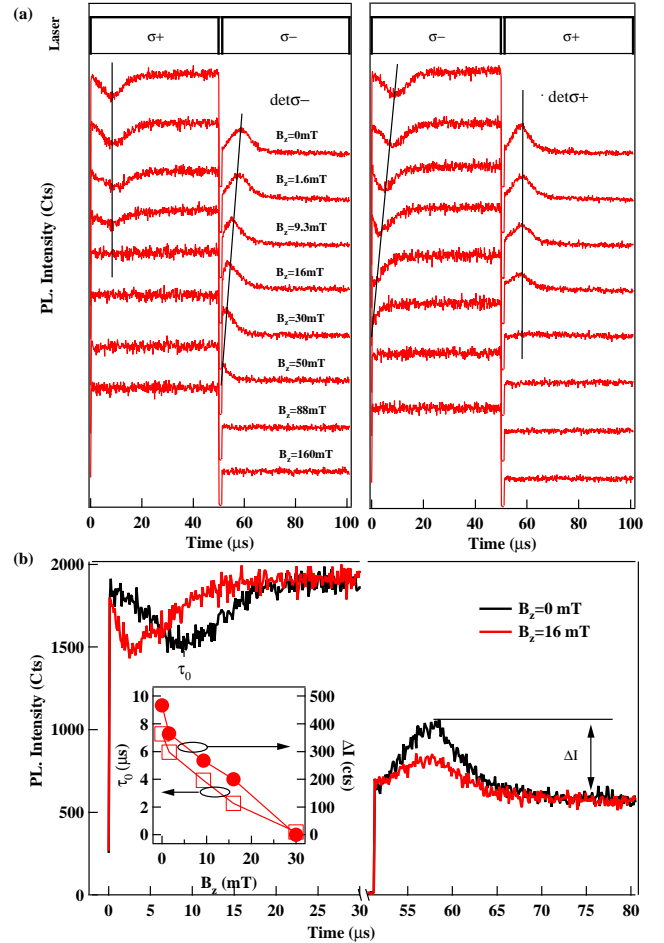}
\caption{(a) Magnetic field dependence of the destruction
and build-up of the nuclear spin polarisation observed
under $\sigma+$/$\sigma-$ modulated excitation (left:
detection $\sigma-$, right: detection $\sigma+$). A
detailed vue of the transients is presented in (b). The
inset shows the magnetic field dependence of the transient
amplitude ($\Delta$I) and position
($\tau_0$).}\label{fig10}
\end{figure}

We observe in Fig.~\ref{fig9}(d), a linear increase of the
pumping rate with the excitation power. While the build up
of DNSP takes a few $ms$ in III-V materials at $B=0T$
\cite{Maletinsky2007}, it occurs in the $\mu s$ range in
our case. This results from the strong localization of the
electron in II-VI quantum dots: the built-up rate of DNSP
scales as\cite{Merkulov2010} $\left|\Psi_e\right|^4$ so
that we typically expect
$\tau_{II-VI}/\tau_{III-V}\approx8000^2/(10^5)^2\approx5.10^{-3}$.

The amplitude of the pumping transient ($\Delta I/I$)
presented in Fig.~\ref{fig9}(c) increases linearly at low
excitation power, reaches a maximum and decreases at high
power. The increase is attributed to an increase of the
nuclear spin polarization. The reduction at high excitation power
likely comes from a decoupling of the dynamics of the
electron spin from the fluctuating nuclear spin. As we have
already seen in Fig.~\ref{fig5}(c), at high excitation intensity,
optical pumping of the electron spin becomes faster
than the precession in the fluctuating field of the nuclear
spin B$_f$ and the measured polarisation rate of the $X^-$
becomes less sensitive to the polarization of the nuclei. A
similar decrease of the amplitude of the transient is
observed under a magnetic field of a few mT applied along
the QD growth axis.

The magnetic field dependence, of $\Delta I/I$ is shown in
Fig.~\ref{fig9}(b). We observe an important decrease of
$\Delta I/I$ as soon as a few $mT$ are applied along the QD
growth axis $z$. This fast decrease mainly comes from the
increase of the relaxation time of the DNSP under magnetic
field (this increase of the relaxation time is evidenced in
Fig.~\ref{fig11} and will be further discussed): As the
nuclear spin polarization does not fully relax during the
dark time, the amplitude of the pumping transient
decreases. This increase of the relaxation time explains
the general shape at fields lower than a few mT. At larger
fields, a decrease also occurs when the static magnetic
field exceed the fluctuating nuclear field. The magnetic
field dependence of $\Delta I/I$ presents an asymmetry as
the magnetic field is reversed. Similarly to the calculated
asymmetry presented in Fig.~\ref{fig0} and the observed
asymmetry presented in Fig.~\ref{fig6}(a), this is the
signature of the creation of an effective internal field
with well defined direction. The faster drop of $\Delta
I/I$  in a positive magnetic field comes from the increase
of the influence of the fluctuating nuclear field B$_f$
when the external magnetic field compensates the Overhauser
field. Such behavior has already been observed on ensemble
of negatively charged CdSe/ZnSe QDs \cite{Akimov2006}.

Our study of the DNSP build-up time scales was complemented
by adding a magnetic field in the Faraday configuration in
the time resolved pumping experiments. Under
$\sigma+$/$\sigma-$ modulated excitation, each switching of
the polarization results in an instantaneous (on the time-scale
of the nuclear spin dynamics) change in the electron
spin polarization followed by a slow evolution due to the
re-polarization of the nuclei. This re-polarization process
is responsible for the minimum observed in the time
evolution of the negative polarization rate. Under a
magnetic field, an asymmetry between the cases of $\sigma+$
and $\sigma-$ excitation is observed in the dynamics of the
coupled electron-nuclei spin system (Fig.~\ref{fig10}).
Under $\sigma+$ excitation, as expected, the application of
a magnetic field along $z$ progressively decreases the
influence of the nuclear spin fluctuations on the
electron-spin dynamics and the minimum in the electron
polarization rate vanishes.

The behavior of the electron polarization is different
under $\sigma-$ excitation: we observe an acceleration with
the increase of B$_z$ of the destruction of the DNSP at the
beginning of the $\sigma-$ pulse. This is illustrated in
the inset of Fig.~\ref{fig11}(b): The position of the
minimum of polarization, $\tau_0$, linearly shift from
$\tau_0\approx 8\mu s$ at $B_z=0mT$ to $\tau_0\approx 0\mu
s$ at $B_z\approx30mT$.

At the end of the $\sigma+$ light train, the polarized
light has created a nuclear field B$_N^{\sigma+}$
anti-parallel to B$_{z}$. At some time after switching to
$\sigma-$ excitation, B$_N$ decreases and approach
-B$_{z}$. At this point, the non-linear feed-back process
in the electron-nuclei "flip-flops" starts and accelerates
the depolarization until the DNSP vanishes. Simultaneously,
the absolute value of the negative polarization reaches a
minimum. Then the nuclei are re-polarized by the $\sigma-$
excitation until B$_N$ reaches B$_N^{\sigma-}$ parallel to
B$_{z}$. Consequently, as observed in Fig.~\ref{fig10}(b),
under $\sigma-$ excitation the destruction of the DNSP is
expected to be faster than its buildup. However, it is not
clear why such a magnetic field dependent acceleration is
not observed during the build-up of the DNSP at the end of
the transient in $\sigma+$ polarization when
B$_N^{\sigma+}$ reaches -B$_{z}$. To fully understand this
behavior, a complete model of the coherent dynamics of
coupled electron and nuclear spins in a weak Faraday
magnetic field should be developed. Such model at zero
field has already shown that the minimum of $\langle
S_z\rangle$ can apparently be shifted from the point
$\langle I_z\rangle$=0 \cite{Petrov2009}.

\subsection{Nuclear spin polarization decay}

\begin{figure}[hbt]
\includegraphics[width=3.5in]{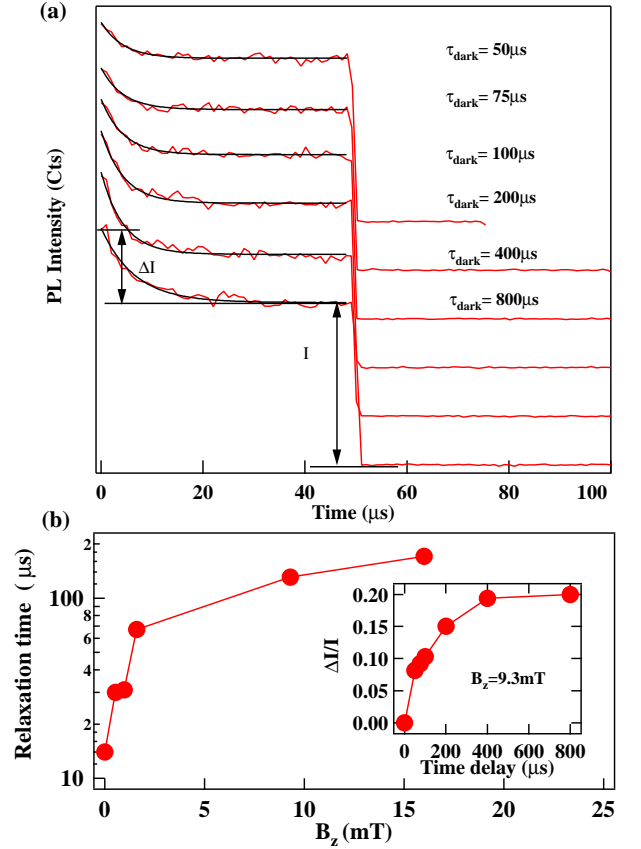}
\caption{(a) Evolution of the DNSP transient, under a
magnetic field B$_z$=9.3mT, with the variation of the dark
time introduced between circularly polarized light trains
of constant length. (b) Magnetic field dependence of the
nuclear spin relaxation time. The inset show the evolution
of amplitude of the DNSP transient with the dark time for
B$_z$=9.3mT.}\label{fig11}
\end{figure}

In order to investigate the variation with magnetic field
of the relaxation time in the dark of the DNSP, we follow
the protocol shown in Fig.\ref{fig11}. For a given magnetic
field, we prepare a DNSP and measure after a time
$\tau_{dark}$ the amplitude of the transient, corresponding
to the partial relaxation of the nuclear polarization. As
$\tau_{dark}$ is increased, this amplitude saturates,
demonstrating the full relaxation.  The variation of the
amplitude of the transient with $\tau_{dark}$ is used to
estimate the relaxation time of the DNSP at a given
magnetic field.

The evolution of this relaxation time is presented in
Fig.\ref{fig11}(b). It ranges from $14\mu s$ at $B=0T$ to
$170\mu s$ at $B=16mT$. The relaxation rate is one order of
magnitude faster than the one observed by Feng {\it et al.}
on ensemble of CdSe/ZnSe QDs\cite{Feng2007}, and the one
expected from nuclear dipole-dipole interactions.
Furthermore, in InAs/GaAs Schottky structure the decay in
electron charged dots occurred in a millisecond time scale
\cite{Maletinsky2007} while nuclear spin lifetime in an
empty dot has been shown to exceed 1 hour.
\cite{Maletinsky2009}.

The magnetic field dependence of the DNSP relaxation
presents a significant increase of the decay time over the
first few $mT$. It has been demonstrated that a magnetic
field of $1mT$ efficiently inhibits nuclear dipole-dipole
interactions in III-V materials \cite{Maletinsky2007}.
Since this interaction is expected to be smaller in our
system with diluted nuclear spins, we can definitely rule
out dipole-dipole interaction as a major cause of DNSP
relaxation.

Co-tunneling to the close-by reservoir could be responsible
for this depolarization. Via hyperfine-mediated flip-flop,
the randomization of the electron spin creates an efficient
relaxation of the nuclei. Following Merkulov {\it et. al.}
\cite{Merkulov2010}, this relaxation time is given by:

\begin{eqnarray}
T_{1e}^{-1}=\frac{2\left\langle
\omega^2\right\rangle\left\|s\right\|^2
\tau_c}{3[1+(\Omega\tau_e)^2]}
\end{eqnarray}

In this expression, $\omega$ is the precession frequency of
the nuclei in the Knight field, $\tau_e$ is the correlation
time of the electron (in the dark), $\Omega$ is the
precession frequency of the electron in the Overhauser
field. At last, $\left\|s\right\|^2$ is equal to
$s(s+1)=3/4$. The fastest relaxation we expect from this
process can be estimated taking
$\Omega=\Omega_{fluc}=2\pi/2ns^{-1}$ and $\tau_c=10ns$. We
obtain $T_{1e}\approx200\mu s$ which is not fast enough.
Therefore, we are tempted to conclude that co-tunneling
alone cannot explain the observed dynamics.

Another mechanism to consider is the depolarization
resulting from an electron-mediated nuclear  dipole-dipole
interaction. This results in exchange constants between the
nuclei which typically scale as $A^2/(N^2\epsilon_z)$. The
resulting rate of nuclear-spin depolarization is
$T^{-1}_{ind}\approx A^2/(N^{3/2}\hbar \epsilon_z)$, where
$\epsilon_z$ is the Zeeman splitting of the electron. This
mechanism could explain a depolarization of the nuclei on a
$\mu s$ scale \cite{Klauser2006}. However, this expression
gives only a minor bound to the relaxation time because the
inhomogeneity of the Knight field can strongly inhibit this
decay \cite{Deng2005, Chekhovich2010}. A magnetic field
along the $z$ axis is expected to affect this process,
progressively decoupling the nuclei from the indirect
coupling created by the electron, as observed in our
experiments in the first few $mT$ (Fig.\ref{fig11}). The
electron-induced nuclear depolarization was demonstrated in
\cite{Maletinsky2007} in which the $ms$ relaxation was
completely suppressed using a voltage pulse on a Schottky
diode in order to remove the resident electron.

\section{conclusion}

In summary, we have evidenced in the PLE spectra of a
negatively charged CdTe QD the polarized fine structure of
the triplet states of the charged exciton. We have studied,
using PL decay measurements, the dynamics of the injection
of spin polarized photo-carriers as a function of the
energy of the injection. We have shown that the injection
above the triplet states of the charged exciton can be use
to pump the resident electron on a time-scale of $10-100ns$
and to create a dynamic nuclear spin polarization. At
$B=0T$, the creation of the dynamic nuclear spin
polarization can be as fast as a few $\mu s$, and the decay
of the nuclear polarization, attributed to an electron
mediated relaxation, is $\approx10\mu s$. The measured
dynamics are $\approx10^3$ faster than the ones observed in
III/V QDs at $B=0T$. The relaxation time of the coupled
electron-nuclei system is increased by one order of
magnitude under a magnetic field of 5mT. The magnetic-field
dependence of the PL polarization rate revealed that the
nuclear spin fluctuations are the dominant process in the
dephasing of the resident electron. We proved that this
dephasing is efficiently suppressed by a large dynamic
nuclear spin polarization at $B=0T$.

\begin{acknowledgements}
This work is supported by the French ANR contract QuAMOS
and EU ITN project Spin-Optronics.
\end{acknowledgements}

\end{document}